\documentclass[aps,pre,twocolumn,groupedaddress,showpacs,superscriptaddress,showkeys]{revtex4}

\usepackage{amsmath}
\usepackage{amssymb}
\usepackage{graphicx}
\usepackage{color}

\begin{document}

\title{Vacancy diffusion in colloidal crystals as determined by
  dynamical density functional theory and the phase field crystal
  model}

\author{Sven van Teeffelen}
\email[]{sven@princeton.edu}
 \affiliation{Department of Molecular Biology, Princeton University,
   Princeton, NJ 08544, U.S.A.}
\author{Cristian Vasile Achim}
\email[]{cristian.v.achim@gmail.com}
 \affiliation{Institut f\"ur Theoretische Physik II, Weiche Materie,
Heinrich-Heine-Universit\"at D\"usseldorf,
D-40225 D\"usseldorf, Germany}
\author{Hartmut L\"owen}
\email[]{hlowen@thphy.uni-duesseldorf.de}
 \affiliation{Institut f\"ur Theoretische Physik II, Weiche Materie,
Heinrich-Heine-Universit\"at D\"usseldorf,
D-40225 D\"usseldorf, Germany}
\date{\today}

\begin{abstract}
  A two-dimensional crystal of repulsive dipolar particles is studied
  in the vicinity of its melting transition by using Brownian dynamics
  computer simulation, dynamical density functional theory and
  phase-field crystal modelling. A vacancy is created by taking out a
  particle from an equilibrated crystal and the relaxation dynamics of
  the vacancy is followed by monitoring the time-dependent
  one-particle density. We find that the vacancy is quickly filled up
  by diffusive hopping of neighbouring particles towards the vacancy
  center. We examine the temperature dependence of the
  diffusion constant and find that it decreases with decreasing
  temperature in the simulations. This trend is reproduced by the
  dynamical density functional theory. Conversely, the phase field
  crystal calculations predict the opposite trend. Therefore, the
  phase-field model needs a temperature-dependent expression for the
  mobility to predict trends correctly.
\end{abstract}
%

%
\maketitle
%
%
\section{Introduction}
Most of the mechanical properties of crystals depend crucially on the
presence of crystalline defects. For material processing it is
therefore of principal importance to understand and control the defect
concentration and dynamics.  The nature and dynamics of defects is
much easier to classify for crystalline sheets in two spatial
dimensions. In this case, it is known for a long time, that the
formation and unbinding of topological defects provides an efficient
way of melting according to the two-stage scenario proposed by
Kosterlitz-Thouless-Nelson-Halperin-Young
(KTNHY) \cite{Strandburg:88}.  Defects can also be accumulated near
edges of crystalline sheets and do occur for two-dimensional crystals
on more complicated manifolds \cite{Nelson:10,Irvine:10}.

Defects are highly dynamic: Whereas the structure of a crystal is
static over long time scales, defects undergo diffusion in the
crystalline background.  The diffusive dynamics of individual point
defects were observed directly in two-dimensional colloidal
suspensions of charged micro-spheres by video
microscopy \cite{Pertsinidis:01, Pertsinidis:05} and they were confirmed and further
analyzed by computer simulations \cite{Libal:07}.  The dynamics of
defects were also explored by real-space methods in a two-dimensional
crystal of weakly damped dust particles in a plasma by Nosenko and
coworkers \cite{Nosenko:11, Nosenko:09, Nosenko:07}, see also \cite{Book}.

Describing the defect dynamics by a microscopic theory is still a
formidable challenge, in particular close to melting. Such a theory
should contain the solid and fluid phase and give a reliable picture
on the defect concentration and its dynamics. Recent progress in this
respect has been made by classical density functional approaches of
freezing \cite{Evans:79,Singh:91,Loewen:94,EmmerichLowen2012AdvPhy}.
For Brownian particles, the density functional approach can be
generalized towards dynamics \cite{Marconi:99,Archer:04,EspanolL2009}
and the dynamics of solidification has been examined in two dimensions
\cite{SvenPRL, SvenRainerPFC}. Similar in line, a more coarse-grained
phase-field-crystal model has been proposed to describe crystal growth
\cite{ElderKHG2002, ElderG2004, ElderPBSG2007, Emmerich2009} and the
defect structure and dynamics, for various applications see
e.g. \cite{YuLV2009,
  JaatinenAEAN2010,AchimRKEGNY2009,RamosAEYN2009,AchimKGYN2006,
  JaatinenAEAN2009, TegzeGTPJANP2009, HuangE2008, MellenthinKP2008,
  McKennaGV2009, WuV2009}.  However a systematic exploration of defect
dynamics by such a density functional theory has not yet been
performed nor has the reliability of the phase-field-crystal model
been systematically checked as far as the trends of defect dynamics
are concerned.

Here, we study the dynamics of vacancies in a two-dimensional
colloidal crystal by using Brownian dynamics computer simulations,
dynamical density functional theory and the phase-field crystal
approach and thereby test the ability of the theoretical approaches to
qualitatively reproduce the observations made in the simulations. The
model system we use here is a two-dimensional suspension of dipolar
colloids. This system has been realized experimentally as
superparamagnetic particles at an air-water interface \cite{Maret:09}.
When exposing superparamagnetic particles to an external magnetic
field perpendicular to the plane, their induced magnetic dipole moment
leads to an effective repulsive interaction whose amplitude can be
tuned by the magnetic field strength.  At sufficiently high field
strength, the system crystallizes into a two-dimensional triangular (i.e.\ hexagonal)
crystal.  This system has been studied extensively by computer
simulations and by the aforementioned dynamical density functional
theory \cite{SvenPRL} and phase field crystal
models \cite{SvenRainerPFC}.

Out of a perfect triangular crystal, some particles are removed and
the relaxation of the resulting defect and its mobility are extracted
by monitoring the one-particle density as a function of time. We
confirm by simulation that the defect mobility is increasing with
increasing temperature, as was already observed for charged particles
by L{\'i}bal {\it et al} \cite{Libal:07}.  This behavior is
qualitatively and semi-quantitatively reproduced in dynamical density
functional theory based on a static Ramakrishnan-Yussouff-like density
functional \cite{Rogers:84, Teeffelen:06}.  The phase-field-crystal
model, on the other hand, fails to predict the trend of the
temperature-dependence of the mobilities.  This is mainly attributed
to the constant kinetic prefactor involved in the phase-field-crystal
approach. For predicting the trend, a temperature-dependent 
corrective mobility is needed for the phase-field-crystal model.

This paper is organized as follows: in section II, we briefly propose
the different approaches used in this paper. Results are
presented in section III and we conclude in section IV.
\section{Theoretical models}
Dynamical density functional theory  and the more coarse-grained
phase-field-crystal model  describe the overdamped Brownian
dynamics in terms of a continuity equation for the deterministic,
time-dependent, and ensemble averaged one-particle density $\rho({\bf
  r},t)$.  Note that in many applications of the phase-field-crystal
model $\rho({\bf r},t)$ is interpreted as a fluctuating density field
that changes for different realizations of the dynamical evolution
even under the same initial conditions.  Here, we will not take this
approach but regard $\rho({\bf r},t)$ as a purely deterministic
quantity. For a more thorough discussion of the physical
interpretation of $\rho({\bf r},t)$ in the phase-field-crystal model
see \cite{SvenRainerPFC,Rauscher}.

The temporal evolution of the density field according to dynamical
density functional theory \cite{Marconi:99,Marconi:00} is given by
\begin{equation}\label{eq:ddft}
  \dot\rho({\bf r},t)
  =\frac{D}{k_BT}\nabla \cdot \left[\rho({\bf r},t)\nabla
\frac{\delta F\left[\rho({\bf r},t)\right]}{\delta \rho({\bf r},t)}\right]\,,
\end{equation}
with $D$ the single-particle diffusion constant and $k_BT$ the thermal
energy. The Helmholtz free energy functional $F[\rho({\bf r})]$ is
provided by classical density functional theory \cite{Evans:79}. In the crystal, the driving current $\rho \nabla(\delta F/\delta \rho)$  obviously assumes the hexagonal symmetry of the underlying crystal. In the interstitial regions, this current is small since the density itself is small.

Note that Eq. (\ref{eq:ddft}) can be derived from first
principles \cite{Marconi:99,Marconi:00}, i.e., from the microscopic
Langevin equations of motion or from the Smoluchowski equation for the
time evolution of their respective probability distribution (for a
review see \cite{SvenRainerPFC}).  Here, we employ the same
approximation to the density functional theory of Ramakrishnan and
Yussouff \cite{Ramakrishnan:79} already introduced in
Refs. \cite{SvenPRL,SvenRainerPFC}. Eq. (\ref{eq:ddft}) then reads
\begin{align}\begin{split}
 \label{eq:ddft2}
\dot\rho({\bf r},t) =
D  \bigg\{ \nabla^2 \rho({\bf r},t)
+(k_BT)^{-1}\nabla\cdot\left[\rho({\bf r},t)\nabla V({\bf r},t)\right] \\
 -\nabla \cdot\left[ \rho({\bf r},t)
   \nabla \int {\rm d}{\bf r}' \rho({\bf r}')
 c_0^{(2)}(|{\bf r}-{\bf r}'|; \rho) \right]\bigg\},
\end{split}\end{align}
where $V({\bf r},t)$ is the time-dependent external
potential. $F_{\rm{ex}}(\rho)$ and $c_0^{(2)}({\bf r};\rho)$ are the
excess free energy and the direct correlation function of the
reference fluid of density $\rho$, respectively.

In this work we consider a two-dimensional system of magnetic dipoles
that are oriented perpendicular to the 2D plane.  The pair potential
of two dipoles in the plane is given by
\begin{equation}
u(r)=u_0/r^3\,,
\end{equation}
where $u_0$ is the interaction strength.  The thermodynamics and
structure depend only on one dimensionless coupling parameter $\Gamma
= u_0 \rho^{3/2}/k_BT$, where $\rho$ is the average one-particle
density and $k_BT$ is the thermal energy.

The two-particle direct correlation function of the fluid
$c_0^{(2)}({\bf r})$ \cite{hansen-mcdonald:86} has been obtained for a
large range of coupling constants $0<\Gamma\leq62.5$ from liquid-state
integral equation theory as previously described \cite{SvenJPCM,
  Teeffelen:06, SvenRainerPFC}.

In order to measure the diffusion of defects, Eq. \eqref{eq:ddft} is
numerically solved on a rectangular periodic box of a fine grid with
$\sim64$ grid points per nearest-neighbor distance $a$.  A finite
difference method with variable time step is applied. The convolution
integrals are solved using the method of fast Fourier transform.

For the more coarse-grained phase-field-crystal model we employ the
two different versions termed PFC1 and PFC2 in
Ref. \cite{SvenRainerPFC}. The PFC1 model constitutes an
approximation of dynamical density functional theory, as introduced
above. The last term in Eq. (\ref{eq:ddft2}) is replaced by its gradient
expansion. The dynamical equation then reads
\begin{align}\begin{split}\label{eq:PFC1}
\dot\rho({\bf r},t)
  =D\nabla^2 \rho({\bf r},t) +
D\nabla \cdot  \Bigg\{ \rho({\bf r},t)
\nabla \Big[ (k_BT)^{-1}V({\bf r},t) \\
-\alpha\left( \hat{C}_0-\hat{C}_2 \nabla^2 + \hat{C}_4 \nabla^4 \right)
\rho({\bf r},t) \Big]\Bigg\}\,.
\end{split}\end{align}
The parameters $\hat{C}_0$, $\hat{C}_2$, and $\hat{C}_4$ are the fit
parameters of a parabola to the first maximum of the Fourier transform
of the two-particle correlation function $\hat c_0^{(2)}(k)$. The
coefficient $\alpha=1.15$ is artificially introduced to match the
melting points of PFC and DDFT.

The second, more coarse-grained model termed PFC2 in
Ref. \cite{SvenRainerPFC}, which is frequently used in the
phase-field-crystal literature, can be obtained from dynamical density
functional theory by assuming a constant mobility, $\rho({\bf
  r},t)=\rho$ in front of the functional derivative in
Eq.~(\ref{eq:ddft}) and a gradient expansion. The model equation then
reads
\begin{align}\begin{split}\label{eq:PFC2}
    \dot\phi({\bf r},t) =D \rho \nabla^2 \Bigg[ \phi({\bf r},t)
    - \frac{1}{2}\phi({\bf
      r},t)^2 +  \frac{1}{3} \phi({\bf r},t)^3 \\
    +(k_BT)^{-1}V({\bf r},t) - \alpha\rho \left(\hat{C}_0-\hat{C}_2 \nabla^2
      + \hat{C}_4 \nabla^4 \right) \phi({\bf r},t) \Bigg] \,.
\end{split}\end{align}
with $\phi({\bf r},t)=[\rho({\bf r},t)-\rho]/\rho$ the dimensionless
density modulation.

The equation of motion is solved using the finite difference method
with a semi-implicit time integration \cite{Chen1999}.

\section{Setup and results}
In the following subsections we qualitatively compare the temperature
dependence of defect diffusion as obtained by computational Brownian
dynamics simulations and as predicted by the dynamical density
functional theory and the phase-field-crystal model 2 (PFC2).
\subsection{Brownian dynamics computer simulation}
\begin{figure}\begin{center}
  \includegraphics[width=8.5cm]{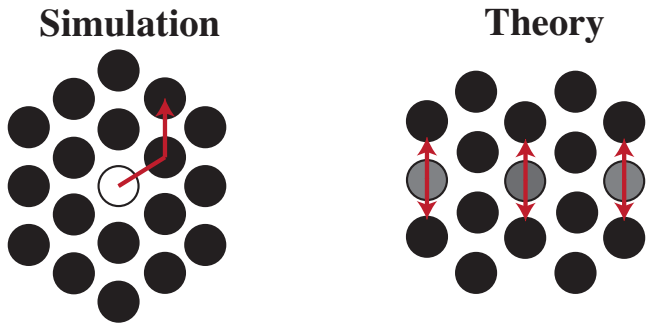}
  \caption{Sketch of the setups in the simulation (left) and theory
    (right): In the computer simulation we remove one particle out of
    a perfect hexagonal lattice and follow the position of the vacancy
    over time.  In dynamical density functional theory and the
    phase-field crystal model we study a quasi-one-dimensional
    relaxation of a depleted central density peaks (grey) being
    replenished by influx of probability density from surrounding
    density peaks.}
  \label{fig:sketch}
\end{center}\end{figure}

\begin{figure}
\begin{center}
  \includegraphics[width=80mm,clip=true]{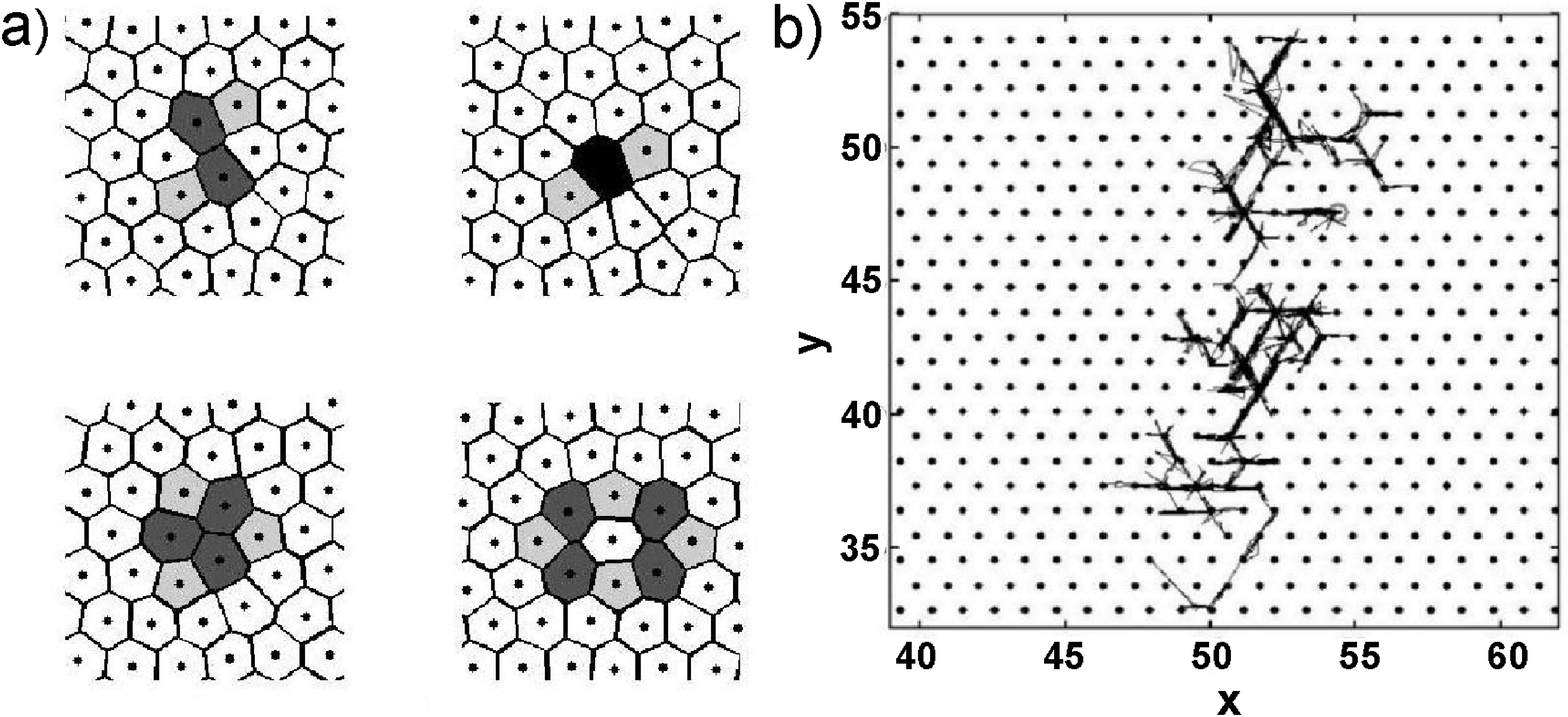}\caption{
    a) A vacancy typically appears as one of four different
    configurations of particles with 5, 7 or 8 neighbors, indicated in
    light, dark grey and black, respectively. b) A typical trajectory of a
    vacancy (similar to results presented in
    Ref. \cite{Libal:07}).}\label{fig:bdconfigurations_trajectory}
\end{center}
\end{figure}

As a reference for the theoretical models we use Brownian dynamics
computer simulations \cite{Ermak:75} to quantify the
diffusion of vacancies for different coupling strengths $\Gamma$ (well
above the melting point at $\Gamma_{\rm
  m}\approx12$\cite{Haghgooie:05}).  Following L{\'i}bal {\it et
  al.} \cite{Libal:07}, we equilibrate a perfectly hexagonal crystal
of $N=2500$ particles in a rectangular, almost square box employing
periodic boundary conditions while keeping one particle tagged at the
origin (see Fig. \ref{fig:sketch}).
Subsequently, the particle at the origin is removed and the diffusion
of the defect is followed over time. In this setup the vacancy concentration is typically relatively small (of the order $10^{-3}$). The position of the defect is determined by a Voronoi construction: The vacancy appears as one of
several configurations of multiple particles with more or less
than six neighbors (see Fig. \ref{fig:bdconfigurations_trajectory}).
The center of mass of these particles is considered as the position of
the vacancy.

\begin{figure}
\begin{center}
  \includegraphics[width=80mm,clip=true]{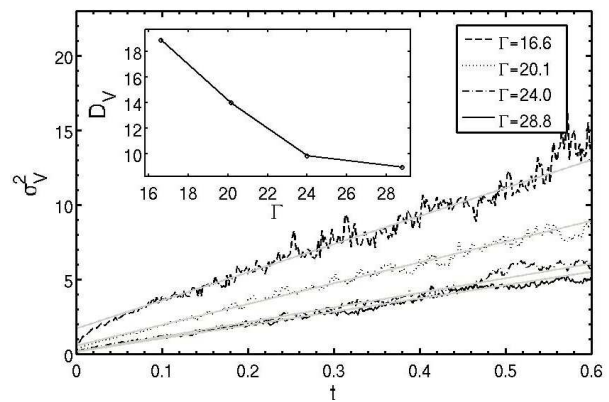}\caption{
    The mean square displacement for $\Gamma=16.6$ (black dashed
    line), $\Gamma=21.12$ (black dotted line), $\Gamma=24$ (black
    dash-dotted line), and $\Gamma=28.8$ (black continuous line)
    obtained from Brownian Dynamics computer simulations. The grey
    continuous lines are linear fits. Inset: the vacancy diffusion
    constant $D_{\rm V}$, calculated from the slope of the mean square
    displacement, as a function of $\Gamma$.  }\label{fig:BDmsqddv}
\end{center}
\end{figure}

As was already observed for Yukawa particles in 2D by L{\'i}bal {\it
  et al.} \cite{Libal:07}, we also find that the defect undergoes
diffusion and that the diffusion constant increases with increasing
temperature, corresponding to a decreasing coupling strength $\Gamma$
(Fig. \ref{fig:BDmsqddv}).  The diffusion constant of the vacancy
$D_{\rm V}$ ranges between $18.9(\rho\tau_{\rm B})^{-1}$, for $\Gamma=16.6$,
and $9(\rho\tau_{\rm B})^{-1}$, for $\Gamma=28.8$.

\subsection{Theory}
In  dynamical density functional theory and in the
phase-field-crystal models, crystals appear as strongly modulated
density fields that have the symmetry of the corresponding
crystal \cite{SvenPRL,SvenRainerPFC}.  These density fields are
mechanically and thermodynamically stable at low temperature or high
coupling strength \cite{Teeffelen:06}.  In an equilibrium density field, the
integrated density field over one Wigner-Seitz cell is equal to the
probability to find a particle at the corresponding lattice site. In
the approximation to the density functional theory by Ramakrishnan and
Yussouff for magnetic dipoles in 2D described above this number is
very close to $1$.  A number smaller than $1$ can be interpreted as a
finite probability to find a vacancy, a number larger than one as a
finite probability to find an interstitial, respectively.

Whereas we have addressed the short-time relaxation dynamics of
crystals in a previous paper \cite{SvenPRL}, we are here concerned
with the long-time dynamics of vacancies.  For ease of computation and
to assess larger time scales we thus start with a slightly different
setup, as compared to the setup in the computer simulations: Instead
of introducing a vacancy with probability $1$ at the center of a large
two-dimensional crystal, which constitutes a problem of cylindrical
symmetry, we study the relaxation dynamics of the
quasi-one-dimensional setup sketched in Fig. \ref{fig:sketch}: Our
reference state
is an equilibrium crystalline density field in a periodic rectangular
box of dimensions $L_x\times L_y=2a\times 64(\sqrt{3}/2)a$, where $a$
is the equilibrium lattice constant.  At time $t=0$ we reduce the
integrated density of all density peaks lying on an infinite row of
neighboring crystal sites along the $x$ axis by a small amount of only $3$ percent, thus rendering the problem quasi-one-dimensional (see Fig. \ref{fig:sketch}). Specifically, low-amplitude Gaussians with the same center positions and similar width as the equilibrated density peaks were subtracted from the density field. Thus, the integrated density of each altered peak is smaller by a few percent than its equilibrium value. The temporal behavior of this setup is expected to be qualitatively similar to that of a cylindrical setup at late times, i.e., once the vacancy has diffused a large distance from its initial position \footnote{Ideally one could have started with an initial density profile which describes an ideal single vacancy, but this causes strong numerical problems in the ideal-gas entropy term.}. Strictly speaking the  long-time diffusion constant $D$ should be calculated from the self-part of a van-Hove function in the presence of interactions (e.g. by using the test particle approach to the DDFT \cite{HopkinsJCP2010}). For a low vacancy density, however, the collective and self-dynamics are expected to be similar such that we have avoided the significant additional effort to implement the test-particle approach.

\begin{figure}\begin{center}
  \includegraphics[width=8.5cm]{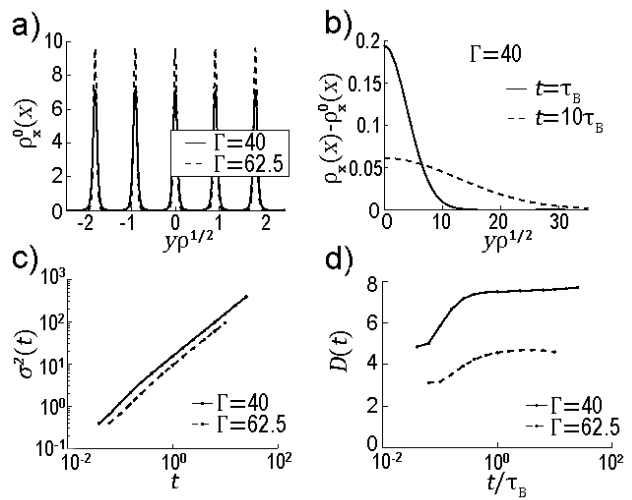}
  \caption{ Results of the dynamical density functional theory: a) the
    initial, equilibrium, $x$-averaged density profiles $\rho_x^0(y,t)$
    for two different values of $\Gamma$, $\Gamma = 40$ and
    $\Gamma=62.5$, the former being close to the melting point at
    $\Gamma=36$, b) the difference of the density profile at time $t$
    and the profile at time $0$ measured at the positions of the peaks
    in (a), such that a set of discrete data set is obtained which is in monotonic in $y$, c) the variance $\sigma^2(t)$ of a Gauss function fitted to
    (b) as a function of time.  d) the effective diffusion constant
    calculated as $D=\sigma_{\rm V}^2/(2t)$.  }
  \label{fig:ddft}
\end{center}\end{figure}
The outcome of the dynamical density functional theory is summarized
in Fig. \ref{fig:ddft} and based on the  $x$-averaged density
field
\begin{equation}
  \label{eq:rhoxavg}
  \rho_x(y,t)\equiv L_x^{-1} \int \mathrm d
 x\,\rho({\bf r},t) \,.
\end{equation}
Fig. \ref{fig:ddft}a) displays the equilibrium averaged density field
$\rho_x(y)=\rho_x(y,t=0)$, i.e., before the introduction of vacancies,
for two different coupling strength, $\Gamma=40$ and $\Gamma=62.5$,
that are close and far from the freezing transition at
$\Gamma_f\approx36$, respectively \footnote {We note that while there are large differences between the freezing point $\Gamma_f$ in DFT and the melting point $\Gamma_m$ in BD, investigating the fluid-solid transition is beyond the scope of this work. Here we simply quenched the system deep enough into the solid state where the details of the equilibrium melting process do not matter much.}. The higher coupling strength
corresponds to higher and more pronounced density peaks.  Removing a
fraction of $0.03$ particles from the row of peaks at $y=0$ leads to a
restoring density current from neighboring particle rows towards the
origin.  This is represented by the difference between the
$x$-averaged density fields of the perturbed and the unperturbed
systems at their original $y$-positions (Fig. \ref{fig:ddft}b):
\begin{equation}
  \label{eq:rhodiff}
  \Delta \rho_x=\rho_x^{0}(y)-\rho_x(y,t)\,.
\end{equation}
For small initial perturbations and at long times the envelope of
$\Delta \rho_x$ approaches a Gaussian function, which broadens over
time.  The variance $\sigma_{\rm V}^2(t)$ is plotted in
Fig. \ref{fig:ddft}c.  As expected, $\sigma_{\rm V}^2(t)$ shows a
linear dependence of $t$ at long times. The long-time slope translates
into a diffusion constant given by $D_{\rm v}=\sigma_{\rm V}^2/(2t)$
(Fig. \ref{fig:ddft}d).  In agreement with the Brownian dynamics
simulations the diffusion constant is higher for low coupling constant
of $\Gamma=40$ ($D_{\rm v}\approx7(\rho\tau_{\rm B})^{-1}$) than for the
high coupling constant of $\Gamma=62.5$ (($D_{\rm
  v}\approx4(\rho\tau_{\rm B})^{-1}$).

\begin{figure}
\begin{center}
  \includegraphics[width=80mm,clip=true]{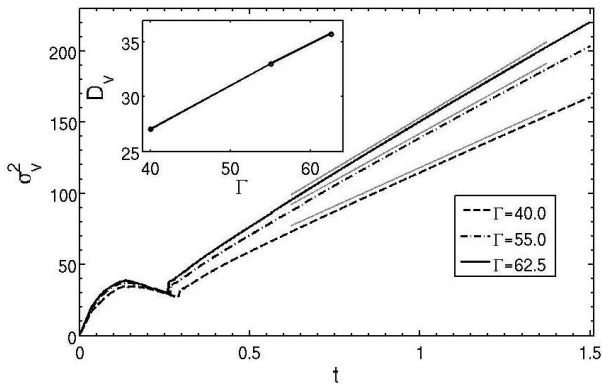}\caption{
    Results of the PFC2 model: The mean square displacement
    $\sigma_{\rm V}^2$ for $\Gamma=40$ (black dashed line),
    $\Gamma=55$ (black dash-dotted line), and $\Gamma=62.5$ (black
    continuous line). The grey continuous lines are linear fits to the
    curves. Inset: the vacancy diffusion constant $D_{\rm V}$,
    calculated from the slope of the mean square displacement as a
    function of $\Gamma$. }\label{fig:PFCmsqddv}
\end{center}
\end{figure}

The same setup is studied in the PFC1 and PFC2 models.
The temporal evolution $\sigma_{\rm V}^2(t)$ of the width of the
Gaussian envelope function describing the relaxing density field is
shown for the PFC2 model in Figure \ref{fig:PFCmsqddv}. After a fast
transient relaxation process its dependence is linear in
time. Remarkably, the corresponding slope $D_{\rm V}$ which is
presented in the inset of Figure \ref{fig:PFCmsqddv} is increasing for
increasing $\Gamma$ conversely to what has been found before in
simulation and DDFT.

The diffusion constant $D_{\rm V}$ increases linearly with $\Gamma$
from $27(\rho\tau_{\rm B})^{-1}$ corresponding to $\Gamma=40$ to
$33(\rho\tau_{\rm B})^{-1}$ and $35.7(\rho\tau_{\rm B})^{-1}$ for $\Gamma$ equal
to $55$ and $62.5$.  The PFC1 model gives the same incorrect trend as
the PFC2 models.  Again, after a transient process
the system reaches a state where the relaxation towards equilibrium is
getting diffusive but the slope increases with increasing $\Gamma$.
The physical reason for the incorrect trend in both variants of PFC
models is that the PFC has a smoothened density profile and therefore
allows for a quick diffusive current of particles from one lattice
site to another. In DDFT, on the other hand, the full density profile
is kept and the density decays to very small values in the
interstitial region between the density peaks. Thus, particle currents
between lattice sites are strongly reduced. For increasing coupling
$\Gamma$, the interstitial density drops even further down and
therefore reduces vacancy diffusion more. This effect is not contained
in both PFC models. Here rather the only remaining trend is set by the
increasing interactions which leads to larger inter-particle forces and
therefore accelerate the dynamics of vacancy diffusion. Therefore, to
account for the correct defect dynamics, the temperature dependence of
the mobility prefactor in the PFC models needs a proper fitting to
reference data.  Though this reduces the predictive power of the PFC
model, it may still be useful for a fast explorative numerical study
for various dynamical effects in solids provided this fitting is
before a priori.

\section{Conclusions}
\label{sec:conclusions}
In conclusion, we have used dynamical density functional theory,
phase-field-crystal theory and particle-resolved Brownian dynamics
computer simulations to calculate the diffusion of defects in a
two-dimensional crystal of repulsive dipoles. The typical diffusion
coefficient of defect motion is expected to decrease with increasing
system temperature as confirmed by the simulation data.  This trend is
reproduced in the dynamical density functional theory but not in the
phase field crystal calculations. These findings show that the PFC
model requires a fitting of the kinetic mobilities as a function of
the thermodynamic parameters if a realistic description of trends is
required to predict material properties. Given the fact that the
efficiency of the PFC model is in general achieved by an optimal
fitting procedure, also for structural predictions
\cite{Oettel_Nestler_et_al_PRE_2012}, such a phenomenological input of
the mobility is an acceptable fact. However, it shows that clearly the
dynamical density functional theory is more appropriate to predict the
microscopic time evolution as a first-principle theory for Brownian
systems.

Future work should address the dynamics of dipolar mixtures of
colloidal particles with different dipole moments
\cite{reviewAdvancesin physicalChemistry} where an equilibrium density
functional for a binary mixture \cite{Kruppa_JCP_2012} is
needed. These mixtures show more complex possibilities of mixed
crystals as a function of the asymmetry in their dipole moments
\cite{AssoudEPL2007}.  In this case one will expect different
diffusion coefficients for different defects topologies. Moreover, one
should do a similar calculation for hard discs, for which a very
accurate functional based on fundamental measure theory
\cite{R._RothJPCMreview2010} was proposed recently
\cite{Oettel_Roth_JCP2012}. A similar comparison can be performed in three dimensions, e.g.\ for hard spheres, 
where the phase field crystal model has been tested against density functional theory recently
\cite{Oettel_Nestler_et_al_PRE_2012}.
Finally it would be interesting to
consider particles with orientational degrees of freedom which form
liquid crystals with interesting defect structures. In fact,
phase-field-crystal models for liquid crystals have been developed
\cite{LowenJPCM2010,WittkowskiPRE2010} and applied
\cite{AchimWittkowski} to freezing recently which opens the way to
describe defect dynamics in liquid crystalline states.
%
\begin{acknowledgments}
 We thank Alexandros Chremos for many inspiring discussions.
 This work has been supported by the DFG through the DFG priority
  program SPP 1296.
\end{acknowledgments}

\end{document}